\documentclass[twocolumn,prl,secnumarabic, nobibnotes, aps,epsfig]{revtex4}

\usepackage{bm}

\def\del#1{{\bf (deleted text)}}

\def\ghost#1{}

\newcommand{\UUNIT}[2]{\,{\rm #1}^{#2}}

\def\dm {DM }

\def\ie {\textit{i.e. }}
\def\cf {\textit{c.f. }}
\def\eg {\textit{e.g. }}
\def\etal {\textit{et al.}}

\def\apj {{ApJ. }}
\def\prl {{PRL }}
\def\pl {{Phys.  Lett. }}

\def\us {\sigma} 
 
\def\uO {\Omega}
 
\def\tHrec {t_{rec}}

\def\vrec {v_{rec}}

\def\mdm {m_{dm}}

\def\beq {\begin{equation}}
\def\eeq {\end{equation}}
\def\dslash#1#2{ \mbox{$#1$ \kern-0.9em \slash \kern0.2em}_{#2} }
\def\thickapprox{\mathrel{%
   \rlap{\raise 0.511ex \hbox{$\sim$}}{\lower 0.511ex \hbox{$=$}}}}

\begin{document}

\title{Can annihilating Dark Matter be lighter than a few GeVs?}

\author{C. Boehm$^1$, T.~A. En{\ss}lin$^2$, J. Silk$^1$}
\affiliation{$^1$ Denys Wilkinson Laboratory, Astrophysics Department, OX1 3RH Oxford, England UK; \\
$^2$Max-Planck-Institut f\"ur Astrophysik  Karl-Schwarzschild-Str. 1, Postfach 13 17, 85741 Garching}

\date{22 August 2002}
\begin{abstract}
We estimate the gamma ray fluxes from the residual annihilations of 
Dark Matter particles having a mass 
$\mdm \in [\UUNIT{MeV}{}, O(\UUNIT{GeV}{})]$ and compare 
them to observations.  
We find that particles lighter than $O(100 \UUNIT{MeV}{})$ are excluded unless 
their cross section is S-wave suppressed. 
\end{abstract}
\pacs{PACS:}

\maketitle

\section*{Introduction}

The accurate measurement of galactic rotation curves, the CMB
spectrum, the primordial abundances of light elements, together with
our understanding of structure formation provide convincing evidence 
in favor of the existence of Dark Matter (DM) \cite{gal}. While the MACHOs
searches \cite{macho} indicate that an astrophysical solution is
rather unlikely, most efforts are now concentrated on searches
for Weakly Interacting Massive Particles (WIMPs) \cite{gunn}.  These
particles would belong to the Cold Dark Matter scenario (CDM); they 
would annihilate and 
suffer from negligible damping effects at a cosmological scale.
Considering fermions only and assuming Fermi interactions, 
it was concluded \cite{lee} that the relic density 
argument constrains the \dm mass 
($m_{dm}$)  to be greater than a few GeVs, as 
is quite naturally predicted within the framework of 
supersymmetry. Nevertheless 
searches for very massive particles ($m_{dm} \agt O$(GeV)) 
remain unsuccessful \cite{cdms}, so there is still 
room for new suggestions.

Alternative \dm scenarios have been proposed in response to
the discrepancy  between observations and CDM numerical
simulations on small scales (which notably predict cuspy haloes 
\cite{swater}). The most ``robust'' one, the Warm Dark Matter scenario (WDM),  
involves \textit{non-annihilating} particles and a very narrow 
range for the \dm mass \cite{joerich}, obtained by requiring
that the free-streaming length be of the order of  
the smallest primordial scale one wants to be compatible with.  
For instance, a typical  warmon  mass  is
$\mdm \sim O(\UUNIT{keV}{}$) for $\sim$ 100 kpc \cite{wdm} but 
this scenario of light and non-annihilating 
DM, although not excluded
as yet, also gives cuspy haloes \cite{devriendt}.

In this letter we propose another model, somehow 
intermediate between  CDM and WDM. 
We shall indeed consider \textit{annihilating} \dm particles having a mass 
$\mdm \in [\UUNIT{MeV}{}, O(\UUNIT{GeV}{})]$. 
This has rarely been studied, probably because of 
the Lee-Weinberg argument that we shall 
evade here by considering bosonic (in fact scalar) 
candidates (for $\mdm \lesssim$ GeV).  
Some of these particles could perhaps 
turn out to be Warm not because of their mass but because of their
collisions with relativistic species \cite{bfs}.  If not excluded by
any cosmological/astrophysical arguments, they could compete with the
collisionless WDM and CDM scenarios. On the other hand, they are likely 
to fail to give flat galactic cores at $\sim 1$ kpc
despite their quite large annihilation rate and a possible significant  
damping mass. 
Interestingly enough, they should escape present \dm direct
detection experiments (which so far are only sensitive to masses
greater than $\sim O(\UUNIT{GeV}{})$), as well as accelerator experiments, 
as briefly discussed in the next section.
These particles would be compatible with the blackbody 
spectrum measurement and will not yield any $^4He$ photodissociation 
(for $m_{dm} > 26$ MeV) provided their (s-wave) cross section satisfies the 
relation 
$ (m_{dm}/\UUNIT{MeV}{}) > 5 \left[\langle\sigma v_r \rangle_{ann}/3 \times 
10^{-27} \UUNIT{cm}{3} \UUNIT{s}{-1}\right] (\Omega_{dm} h^2)^2 $ 
(assuming self-conjugate DM particles, 
$H \!=\!100 h \UUNIT{km}{}/\UUNIT{s}{}/\UUNIT{Mpc}{}$ and 
using the $D$ measurement only \cite{turner}).

Since light particles could yield gamma rays at energies that 
have been already probed experimentally, 
we mainly focus on their indirect detection 
signature to determine whether or not it is reasonable to consider them.  
We find that the gamma ray 
fluxes associated with particles lighter than $O(100 \UUNIT{MeV}{})$ are in 
conflict with observations unless the 
$v^2$-dependent term in the annihilation cross section 
 i) is much larger than 
the S-wave term (developed at the first order) and ii)  
satisfies the relic density requirement. Radio fluxes 
also validate this conclusion.

\section*{Acceptable values of the cross sections}

Relic density calculations provide a strong constraint on any DM candidate.  
When DM particles are able to annihilate (\ie when their non-relativistic 
transition occurs before their thermal decoupling), one obtains  
a simple relationship, independent of $m_{dm}$, between the 
DM cosmological parameter $ \uO_{dm}^{th} h^2$ 
and the total annihilation cross section  
\cite{omega}. 
By requiring that $ \uO_{dm}^{th} h^2 $ matches 
the observed value ($\sim$ 0.1, see \cite{wmap}), 
one gets the following approximate annihilation cross section 
\begin{equation}
\langle \us v_r \rangle_{ann} \simeq 7 \ 10^{-27} \ \frac{x_F}{\sqrt{g_{\star}}} \
 \left(\frac{\Omega_{dm} h^2}{0.1}\right)^{-1}\  \ \UUNIT{cm}{3} \UUNIT{s}{-1}
\label{san}
\end{equation}
with $x_F  \!= \! m_{dm}/T_F \! \simeq 17.2 + \, \ln(g/\sqrt{g_{\star}}) + \, \ln(m_{dm}/\UUNIT{GeV}{}) + \, \ln\sqrt{x_F} \in $[12-19] for particles in the
MeV-$O$(GeV) range 
($g$ and  $g_{\star}$ being the number of internal and relativistic degrees 
of freedom respectively). 
We write the cross section as 
$\us v_r  \sim a + b (v/c)^2$ where $a$ and $b$ are 
some constants related to the S and P-wave terms (here $v$ is the DM velocity   
and $v_r$ the relative velocity). We shall assume first that $a \gtrsim b$ 
(and take $c = 1$). 

Eq.(\ref{san}) sets the maximum value of $\langle \us v_r \rangle_{ann}$ 
one can use to compute  
gamma ray and radio fluxes from DM residual annihilations. 
Larger cross sections are possible
if $n_{dm} \neq \bar{n}_{dm}$ (but  
residual annihilations are unlikely or even impossible 
since no, or few, anti-particles would be left after the \dm freeze-out, 
$t_{fo}$) or if the co-annihilation mechanism (which involves DM and $X$ 
particles) is at work \cite{griest}. In this case, however, the relationship 
$ \frac{m_X}{m_{dm}} \simeq 1 + x_{F}^{-1} \ln\left(\frac{g_X}{g_{dm}} 
\left(\frac{m_X}{m_{dm}}\right)^{3/2} \frac{\sigma_{coan}}{\sigma_{ann}}
\right)$ indicates that $m_X$ should be close to $\mdm$, 
which is actually excluded if $\mdm < O(\UUNIT{GeV}{})$ and if 
$X$ is a charged particle. Thus, unless there exists a neutral 
(long-lived) particle $X$, the maximum 
annihilation cross section (times relative velocity) 
 into ordinary particles (\eg $\gamma, e^-$) 
that is legitimate to consider is about 
$10^{-26} \UUNIT{cm}{3} \UUNIT{s}{-1}.$

\subsection{The least massive annihilating WIMP}

Since eq.(\ref{san}) is almost independent of the DM  mass 
(the only dependence in $\mdm$ being the logarithm which is 
``hidden'' in $x_F$), even light candidates are expected to be 
allowed. However, when dealing with the range $\mdm \alt O(\UUNIT{GeV}{})$, 
having $\langle \sigma v_r \rangle_{ann} \sim 10^{-26} 
\UUNIT{cm}{3} \UUNIT{s}{-1}$ rather favors particles with an annihilation 
cross section independent of the \dm mass or, for which the annihilations  
rely on the exchange of light neutral particles. Taken at face value, 
the first point strongly suggests that light \dm should be made 
of scalars (which is in agreement with the second) and with non chiral 
couplings \cite{bf}. One now has to check that their damping scale 
is not too large.

In contrast with the free-streaming length ($l_{fs}$) of  
non-annihilating DM particles which depends only on $\mdm$,  
the scale $l_{fs}$ of interacting particles in the 
$[\UUNIT{MeV}{}, O(\UUNIT{GeV}{})$] mass range 
depends on both the DM mass and interaction rate \cite{bfs}.
By imposing $l_{fs} \lesssim 100$ kpc (the scale of the  
smallest galaxies) and assuming that interactions are weak enough  
to imply a decoupling in the radiation dominated era, 
one finds that particles with   
$m_{dm} \alt  O(\UUNIT{MeV}{}) \, (a_{dec}/10^{-4})$  
induce a cut-off in the matter power spectrum at 
$\sim 100$ kpc ($a_{dec}$ being the scale-factor at the DM thermal decoupling). 
Nevertheless, in a realistic particle physics model, 
the thermal decoupling (based on the estimate 
of the DM-e elastic scattering cross section) is  seen 
to occur around 1 MeV ($a_{dec} \sim 10^{-10}$), so the mass range 
above 1 MeV should not modify the matter power 
spectrum at cosmological scales. 
In fact, even if the DM thermal decoupling was in the matter 
dominated era, only masses of  
$ m_{dm} \lesssim 15 \, \UUNIT{MeV}{} \, (\Omega_{dm}
h^2/0.1)^{-1} \,$ would affect the $10^9 M_{\odot}$ scale. 
The limit on $\mdm$ would actually get even smaller 
if DM was thermally decoupling after the non-linear collapse 
\cite{bfs} (\cf self-interacting DM for instance \cite{spergel}, 
which now appears unlikely \cite{clsi}), as the primordial fluctuations 
``disappear'' to form objects.   
Collisional damping due to $\nu$-DM interactions may also contribute but 
is expected to yield a damping mass 
$< 10^9 M_{\odot}$ so 
light annihilating candidates certainly deserve to be studied.

We note that 
low DM masses are not constrained by direct detection experiments 
which are only sensitive to masses greater 
than $\sim 7$ GeV \cite{cdms} (except for \eg Cresst \cite{Cresst}, 
MACHe3, ROSEBUD, Tokyo \cite{Rosebud} 
which are or will be able to go as low as 
$\sim$ 1 GeV). 
In fact, exploring the low \dm mass region should be 
a problem for cryogenic detectors 
since the detection mechanism they currently use (based on nucleus 
recoil) does not allow for the detection of particles much 
lighter than $\sim$1 GeV without a  
significant effort (even by using the lightest possible nucleus). 
Light scalars could escape searches in 
$e^+ e^-$ colliders even when their 
production is based on the exchange of massive fermions  
(with a mass $\agt 100$ GeV) as their cross section for   
anomalous single photon events is still below (albeit very close) 
the sensitivity of past experiments \cite{bf}.   
Moreover, if they are able to annihilate into photons, the most interesting 
signature   $e^+ e^- \rightarrow e^+ e^- \dslash{E}{} \dslash{E}{}$ 
($\dslash{E}{}$ denoting the DM particles \ie some missing energy) 
at the $\alpha^4$  $\alpha'^2$
order should also  be 
invisible because the electrons should  remain mainly in the beam pipe,  
as one can infer from a kinematic analysis. Nevertheless, 
even if there exists a deviation large enough to be 
detected, it is unlikely that PETRA or LEP experiments, for instance, 
got enough sensitivity. 
Hence,  particle physics experiments 
still allow for the range we consider.
Note that in any case \dm should not have a coupling to the $Z$ boson 
(otherwise it would have been detected in accelerator experiments) but  
since bino particles or right-handed neutrinos, for instance, do not have this 
coupling either, this assumption seems reasonable.

\section*{Indirect detection during the recombination epoch}

Let us check if \dm annihilations 
at the recombination epoch yield enough
redshifted photons at an energy $E_{\gamma} \in [\UUNIT{keV}{}, 10
\UUNIT{MeV}{}]$ to be detected nowadays.  Assuming $a \gtrsim b$, 
one gets the following gamma-ray number density
$n_{\gamma}^{\rm rec} \approx 2 \langle\sigma v_r \rangle_{ann} \ {(n_{\rm
dm}^{\rm rec})}^{ 2} \ \tHrec$ (with $n_{\rm dm}^{\rm rec}$ and 
$t_{\rm rec}$ the \dm number density and Hubble time at the recombination
epoch). This yields the present day flux
$ 
\Phi  \! \sim \! \!    \, \left(\frac{\Omega_{dm}
h^2}{0.12}\right)^2 \,  \langle \sigma v_r \rangle _{26} \ m_{\rm MeV}^{-2}
\, \UUNIT{cm}{-2} \, \UUNIT{s}{-1} \, \UUNIT{sr}{-1}
$
using  $\Phi \!= \!  \! c \ n_{\gamma} / 4 \pi$, $\langle \sigma v_r\rangle_{26} \!= \!\langle \sigma v_r \rangle_{\rm ann}/\rm 10^{-26} \, cm^3 \, s^{-1}$ and
$ m_{\rm MeV} \!= \! \mdm/(\rm MeV/c^2)$.
P-wave annihilation cross sections would decrease this flux 
\cite{pwaveATanni} but the latter is already overestimated since 
$O$(MeV) photons can lose some energy by scattering on remnant free 
electrons and by reionizing atoms 
\cite{1989ApJ...344..551Z}.
Comparing $\Phi$  with the much higher
observed fluxes, namely 
$\Phi_{obs}^{[1-30] \UUNIT{keV}{}}(>E_{\rm min}) \sim 20\, (E_{\rm
min}/\UUNIT{keV}{})^{-0.4}/(\!\UUNIT{cm}{2} \UUNIT{s}{} \UUNIT{sr}{})$, 
$\Phi_{obs}^{[0.1,10] \UUNIT{MeV}{}}(>E_{\rm min}) \sim 3\
10^{-3}\, (E_{\rm min}/\UUNIT{MeV}{})^{-1.5}/(\!\UUNIT{cm}{2}
\UUNIT{s}{} \UUNIT{sr}{})$ and
$\Phi_{obs} \approx 4\,(1.5)\ 10^{-5}/(\!\UUNIT{cm}{2} \UUNIT{s}{}
\UUNIT{sr}{})$ for the range [30, 100] MeV (and $>100$ MeV
respectively) \cite{1996A&AS..120C.607H}, we find that 
such residual annihilations are not ruled out by recent data.

\section*{Dark Matter haloes}

\begin{table}[ht]
\vspace{-0.5em}
\begin{tabular}{ccccccccc}
\hline
\hline
 & $\alpha$ & $\beta$ & $\gamma$ & $r_{\rm s}$ & 
 \multicolumn{3}{c}{$F(\theta)$} & 
$\!\!\!\!\!\! \Phi /(\langle \sigma v_r
\rangle_{26}\, m_{\rm GeV}^{-2})$\\
& & & & kpc &  
$1^\circ$ & $10^\circ$ & $45^\circ$ & $\UUNIT{cm}{-2}\UUNIT{s}{-1}$\\
\hline
NFW & 
1 & 3 & 1 & 25 & 
0.077 & 0.62 & 1.7 &
$5.9\ 10^{-6}$
\\
KRA & 
2 & 3 & 0.2 & 11 & 
$1.7 \cdot 10^{-4}$ & 0.014 & 0.15&
$7.5\ 10^{-8}$
\\
ISO & 
2 & 2 & 0 & 4 & 
$1.2 \cdot 10^{-4}$ & 0.011 & 0.08&
$1.8\ 10^{-7}$
\\
BE & 
1 & 3 & 0.3 & 4 & 
$1.2 \cdot 10^{-4}$ & 0.004 & 0.01&
$4.1\ 10^{-6}$\\
\hline
\hline
\end{tabular}
\caption{\label{tab:gc} Angular function $F(\theta)$ and central
$\gamma$-ray flux $\Phi(<1.5^\circ)$ for different galactic \dm
profiles, $R_{\rm sol} = 8.5 \UUNIT{kpc}{}$ and $\rho_0$ 
chosen so that $\rho(R_{\rm sol}) = 0.3 \,{\rm GeV/c^2\, cm^{-3}}$
\cite{BergstroemUllio}.}
\vspace{-1em}
\end{table}

Haloes are promising targets for \dm annihilation
radiation searches due to their high central density and small
cosmological distance \cite{BergstroemUllio}. Here, we assume a \dm
distribution of the form:
$
\rho(r) = \rho_0 \, \left(r/r_s \right)^{-\gamma} \, \left[1 +
\left(r/r_s \right)^{\alpha}\right]^{-{(\beta -\gamma)}/\alpha}
$
where $(\alpha,\beta,\gamma)$ are given in Table \ref{tab:gc} for
the Navarro-Frenk-White (NFW), Kravstov \textit{et al.} (KRA),
modified isothermal (ISO) and Binney$\&$Evans (BE) models.  
The number of photons produced per unit of time 
is given by: $\dot{n}_\gamma(r) = \dot{n}_0 \, ({\rho(r)}/{\rho_0})^2$
with $\dot{n}_0 = 2 \,\langle \sigma v_r \rangle_{ann} \
({\rho_0}/{\mdm})^2$.

$\bullet$ The radiative flux from \dm annihilations inside the Milky
Way being distributed over a large fraction of the sky, we can compute the flux 
reaching the Earth from a circular cone with an opening
angle $2\,\theta$ centered on the galactic centre: $\Phi(< \! \theta) =
\dot{n}_0\,r_{\rm s} \,F(\theta) =$ \ $ 6.2 \ 10^{-4}\, {\rm cm^{-2}\,
s^{-1}}$  $\langle \sigma v_r\rangle_{26}\, m_{\rm GeV}^{-2}$ $
({\rho_0}/({{\rm GeV\, cm^{-3}}}))^2 $ \ $ {r_{\rm s}\,F(\theta)}/({{\rm
10\, kpc}})$, 
where $F(\theta)$ is a dimentionless function which depends on the 
\dm profile.  
Numerical values of the flux are listed in Table \ref{tab:gc}. They  have to be 
compared with EGRET data from the galactic centre 
\cite{mayer-hasselwander}: 
 $\Phi_{obs}^{gc}(<1.5^\circ,\, E_\gamma>E_{\rm min}) = 2.2 \ 10^{-6} \UUNIT{cm}{-2} \UUNIT{s}{-1} \ (E_{\rm min}/(100\UUNIT{MeV}{}))^{-0.3}$ 
within [30 MeV, 1 GeV] (and even less 
within [1 GeV, 10 GeV]) which is in agreement with the COMPTEL 
flux  $\Phi_{obs}^{gc}(<1.5^\circ,\, E_\gamma>\UUNIT{MeV}{}) \sim 10^{-5} 
\UUNIT{cm}{-2} \UUNIT{s}{-1}$ (or $\sim 10^{-4}$ within $5^\circ$).

\begin{table}[h] 
\vspace{-0.5em}
\begin{tabular}{llllllll}
\hline
\hline 
& $\alpha$ & $\beta$ & $\gamma$ & $r_{\rm s}$ & $D$ & $\rho_0$ & $\!\!\!\!\!\!\!\!\!\!\! \Phi_{cl}/(\langle \sigma v_r\rangle_{26} m_{\rm GeV}^{-2})$ \\
& & & & kpc & Mpc &  $\UUNIT{GeV}{}/\UUNIT{c}{2}\UUNIT{cm}{3}$ & $\UUNIT{cm}{-2} \UUNIT{s}{-1}$ \\
\hline
C-NFW & 
1 & 3 & 1 & $0.25/h$ & $70/h$ & $0.090 h^{2}$ & $5.3\,10^{-10}h^{3}$\\
C-$\beta$-pr. & 2 & $2.25$ & 0 & $0.2/h$ & $70/h$ &$0.13 h^{2}$ & $8.8\,10^{-10}\,h^{3}$\\
V-NFW & 
1 & 3 & 1 & 0.56 & $15$ & $0.012$ & $2.4\,10^{-9}$\\
V-$\beta$-pr.& 2 & $1.41$ & 0 & 0.015 & $15$ & $0.76$ & $3.0\,10^{-9}$\\
\hline
\hline
\end{tabular}
\caption{\label{tab:cl}Expected fluxes from the Coma (C) and Virgo (V)
cluster for different \dm profiles \cite{cl.profiles}. For the
$\beta$-profile of Virgo, only the flux within 1 Mpc is given. $h =0.7$.}
\vspace{-1em}
\end{table}

$\bullet$ The total gamma ray flux from the \dm halo of a galaxy cluster 
located at distance $D$ is well approximated by 
$\Phi_{\rm cl} = \int\!\! dV \, \dot{n}_\gamma(\vec{r}) /(4\pi\,D^2)$. 
We list it for different 
halo profiles of two nearby clusters (Table \ref{tab:cl}) 
\cite{cl.profiles}.  These values have to be compared with 
$\Phi_{obs}^{cl}(>100 \UUNIT{MeV}{} ) \! \! < \!  4 \, 10^{-8} \UUNIT{cm}{-2} 
\UUNIT{s}{-1}$ at 2 sigma for these two clusters \cite{coma}. 

By comparing the numbers in Table \ref{tab:gc} with $\Phi_{obs}^{gc}$, 
we conclude that the annihilation 
cross section into $2 \gamma$ (times relative velocity) 
of particles lighter than $O(100)$ MeV should be 
much below $\langle \sigma v_r\rangle_{\rm ann} \sim 10^{-26} \UUNIT{cm}{3} 
\UUNIT{s}{-1}$. 
This is clearly in conflict with the relic abundance requirement unless one 
assumes that i) $\sigma v_r$ is dominated by $b v^2$ at the freeze-out epoch 
and ii) the a-term is smaller or equal to 
$10^{-31} - 10^{-28} \UUNIT{cm^3}{} \UUNIT{s}{-1}$  for 
$m_{dm} \in [1, 100]$ MeV respectively (\ie $\lesssim 10^{-5} - 10^{-2}$  
times the b-term at the freeze-out epoch). 
On the other hand, heavy particles annihilating into photons 
and with $m_{dm} \agt O(\UUNIT{GeV}{})$ 
appear quite compatible with observations even when $\sigma v_r \propto const$ 
(regardless of the profile used).

If the \dm residual annihilations mainly proceed 
into leptons, then electrons could be detectable via synchrotron emission
in a magnetic field $B$ \cite{synchrotron}. For the galactic centre 
only, the flux 
$\mathcal{F}_{\nu} (<0.1^\circ) \approx 5861.04 \, {\rm Jy} (\nu/{\rm GHz})^{-0.5}\, m_{\rm GeV}^{-2}\, \langle \sigma v_r\rangle_{26} \,  (B/{\rm mG})^{-0.5}$ is compatible with the observed radio flux of SgrA 
(of the order of 360 Jy at $\nu = 330$ MHz) provided $m_{dm} \gtrsim 10$ GeV 
(assuming that all the annihilations proceed into $e^+ e^-$, a NFW profile 
and $B \sim 1$ mG). 
However, masses $\mdm  \lesssim 10 \UUNIT{GeV}{}$ are OK if 
$\sigma v_r \thickapprox  b v^2\vert_{t_{fo}}$. 
Majorana fermions that would annihilate into $e^+ e^-$ through 
the exchange of scalar particles may actually 
have this property (if they are chirally coupled and 
for some value of $\mdm$) but they would 
fail to give the correct relic density if $\mdm \alt O(\UUNIT{GeV}{})$. 
In contrast, scalars coupled to fermions have  
the correct relic density but not the correct radio flux. 
But these particles satisfy both criteria if   
they exchange a light gauge boson. 
For Coma (Virgo being swamped with 
emission from M87) only $m_{dm}  \agt 10 \UUNIT{GeV}{} $ produces radio
emission at observed frequencies but the latter is 
(depending on $B \in [0.1,10]\,\mu$ G) 
of the same order of magnitude as the observed flux 
for $m_{dm} \sim 10$ GeV only
(it is less for higher masses).

\section*{Discussion and conclusion}

We now discuss three specific mass interval which have not  
been studied significantly before. They represent a ``hole'' in the
investigations of the \dm parameter space and deserve to 
be studied at least to probe whether or not there exist alternatives 
to  heavy DM particles.

\textbf{$m_{dm} \in [1 \UUNIT{MeV}{},m_{\mu}[$}:  Since 
the annihilation cross section of a pair of scalars into $e^+ e^-$  
(via a fermion exchange $F$ and non-chiral couplings) is expected to be 
free of $m_{dm}$ (and potentially of the order of eq.(\ref{san}), see 
\textit{e.g.} \cite{sneutrino}), one can discuss the case of 
light scalar DM candidates.  However, to be compatible with 
the galactic centre COMPTEL/EGRET data, their annihilation 
cross section should be dominated by a term in $\, v^2$ at $t_{fo}$. 
This is actually in conflict with the $F$ exchanges (needed to get the 
correct relic density)  
but one can postulate an asymmetry between the \dm and anti-\dm 
number densities or assume that annihilations mainly proceed 
through the exchange of either 
a light neutral fermionic WIMP or gauge boson.
In any case, a more careful study is needed to ensure 
that the particles introduced satisfy all experimental constraints \cite{bf}.
Such light \dm particles could perhaps be detected or excluded 
by using their interactions with ordinary matter 
or searching  for the particles supposed to be 
exchanged in the annihilation process.

\textbf{$m_{dm} \in [m_{\mu}, m_{\tau}[$}:
They may be compatible with observed fluxes and  relic density 
even for $a \gtrsim b$  
but the production of $D+ ^3He$ 
\cite{turner} tends however to favor an 
annihilation cross section dominated by the term in $v^2$. 

\textbf{$m_{dm} \in[m_{\tau}, O(10 \UUNIT{GeV}{})]$ }: 
They seem in agreement with observations even when 
$a \sim 10^{-26} \UUNIT{cm}{3} \UUNIT{s}{-1}$ 
but, if they mainly annihilate into $e^+ e^-$, 
radio fluxes tend to favor $\sigma v_r \propto b v^2 \vert_{t_{fo}}$. 
If they mainly annihilate into a pair  $\tau-\bar{\tau}$, one expects 
a soft gamma ray emission (notably from $\pi$ decays) 
plus an excess of positrons (due to $\bar{\tau}$  decays).  
Their energy is expected to be  
$E_{e^+} = (E_{\bar{\tau}} - \langle E_{FSR}\rangle) (1 + \sqrt{1-\frac{m_{\bar{\tau}}^2}{E_{\bar{\tau}}^2}} \cos \theta)/2 \in [0, m_{dm}- \langle E_{FSR}\rangle]$, where 
$ \langle E_{FSR}\rangle = E_{\gamma} \simeq
\frac{ 2 \, m_{dm} \, \beta}{\beta+1} (1- m_{\bar{\tau}}/m_{dm})^{\beta+1}$   
is the energy lost 
by the taus  under the form of a single 
photon emission, \ie via the Final State Radiation mechanism (FSR).
As an illustration, with  $\beta \sim 0.09$ and 
$m_{dm} \sim 10 \UUNIT{GeV}{}$, one finds 
$\langle E_{FSR}\rangle \sim \, 1.5 \UUNIT{GeV}{}$ indicating that  
the emitted positrons could have an  
energy of $E_{e^+} \simeq 8.5 \UUNIT{GeV}{}$ which seems 
of the order of the lower bound of the energy range indicated by 
HEAT experiment\cite{heat}. 
Of course, $\mdm \lesssim O(10 \UUNIT{GeV}{})$ will fail to 
explain the $e^+$ excess above $\sim$ 10 GeV 
and more careful studies are required 
to check whether particles with 
$m_{\tau}< \mdm < O(10 \UUNIT{GeV}{})$ can indeed escape present 
\dm searches (which could perhaps be the case if 
they have stronger interactions with $e^-$ than with 
nuclei, giving them a potential signature).

\section*{Acknowledgment}
The authors would like to thank J. Dunkley, 
M. De Jesus, M. Fairbain, J.M. Fr\`ere, 
K. Griest, S. Hansen, P. Janot, K. Jedamzik, C. Jones, 
G. Madejski, P. de Marcillac, K. Moodley, V. Springel, and F. Stoehr.
C.B. is supported by an individual PPARC Fellowship.


\begin{thebibliography}{99}


\bibitem{gal} A. Bosma, IAUS, 100, 11B (1983); 
V.C. Rubin,  IAUS, 100, 3R, (1983); 
J. L. Sievers \textit{et al.}, astro-ph/0205387;
K. A. Olive, astro-ph/0202486;
M. Davis \textit{et al.}, ApJ. 292: 371-394, 1985.

\bibitem{macho} The Macho collaboration, ApJ. Lett. 550, L169 (2001).

\bibitem{gunn} J. E. Gunn \textit{et al.},  ApJ. 223, 1015G (1978).  


\bibitem{lee} B. W. Lee and S. Weinberg, PRL 39, 4, 183 (1977).

\bibitem{cdms} see \eg CDMS collaboration, astro-ph/0203500; 
Edelweiss, astro-ph/0206271; 
DAMA, astro-ph/0205047.  

\bibitem{swater} B.~Moore \textit{et al.}, \apj Lett.\ 524, L19
(1999); R.~A.~Flores and J.~R.~Primack, \apj Lett.\
427, L1 (1994);J.J. Binney and N.W. Evans, astro-ph/0108505.  
The interpretation of rotation curves of Low Surface 
Brightness and dwarf galaxies is however still under debate: 
F. C. van den Bosch, R. A. Swaters, MNRAS, accepted, astro-ph/0006048. 


\bibitem{joerich} R. Schaeffer, J. Silk, ApJ. 292, 319 (1985).  

\bibitem{wdm}  
V. K. Narayanan \textit{et al.}, ApJ. 543L 103N (2000); 

\bibitem{devriendt} A.~Knebe \textit{et al.}, Month.\ Not.\ Roy.\
Astr.\ Soc.\ 329, 813 (2002);
R. Barkana \etal, ApJ. 558, 482B (2001); 

\bibitem{bfs} C.~Boehm \etal , \pl  B 518, 8; 
C. Boehm \textit{et al.}, will appear in Phys.Rev.D [astro-ph/0112522]. 

\bibitem{turner}  P. McDonald \etal, Phys.Rev. D63, 
023001 (2001); J. A. Frieman \etal, Phys.Rev.D 41, 10, (1990).  

\bibitem{omega} M. Drees, M. M. Nojiri, Phys.Rev. D47, 376, (1993). 

\bibitem{wmap}
D. N. Spergel \textit{et al.}, astro-ph/0302209. 

\bibitem{griest} K. Griest, D. Seckel,  Phys. Rev. D 43, 3191, (1991).   

\bibitem{bf} C.~Boehm, P.~Fayet, hep-ph/0305261.  

\bibitem{spergel}  D. N. Spergel, P. J. Steinhardt, \prl 84,  3760, 
(2000). 

\bibitem{clsi} M. W. Bautz \etal, astro-ph/0202338.   
 
\bibitem{Cresst} Cresst, astro-ph/0106314.  

\bibitem{Rosebud} G. Chardin, Les Houches Summer School, (1999). 


\bibitem{pwaveATanni}
Assuming a symmetry between the elastic scattering and the
annihilation cross section, 
one finds $\vrec \agt O(\UUNIT{km}{} \UUNIT{s}{-1})$ reducing
the predicted flux by a factor at most $10^{-10}$.

\bibitem{1989ApJ...344..551Z} One can replace $a_{rec}$
by the scale-factor associated with the moment the photons can escape
to their optical depth ($a_{depth} > a_{rec}$), which also reduces the
flux, see A.~A.~Zdziarski \& R. Svensson, ApJ. 344, 551 (1989).

\bibitem{1996A&AS..120C.607H} G. Hasinger, A\&A Supp.  120,
607 (1996), and references therein; 
P. Sreekumar, \textit{et al.}, ApJ. 494, 523 (1998).

\bibitem{BergstroemUllio} L. Bergstr{\" o}m \etal, 
Astroparticle Physics, 9, 137 (1998), and references therein.


\bibitem{cl.profiles} J.~P. Hughes, Untangling Coma Berenices:
A New Vision of an Old Cluster, 175 (1998); U.~G. Briel \etal, 
A\&A, 259, L31 (1992); McLaughlin, ApJ. 512, L9 (1999);
H. B{\"o}hringer, private communication.


\bibitem{mayer-hasselwander}
H.A. Mayer-Hasselwander,  \textit{et al.}, A\&A 335, 161;
V. Sch{\"o}nfelder, \textit{et al.}   A\&AS, 143, 145 (2000).

\bibitem{coma} P. Sreekumar \textit{et al}, ApJ. 464, 628 (1996).

\bibitem{synchrotron} The calculated gamma ray fluxes
can be translated into radio flux. To a good approximation, $F_\nu = 45
\,{\rm Jy}\,{f}_{B}\, \nu_{\UUNIT{GHz}{}}^{-1/2}\,\Phi_{10}$ for 
frequencies $\nu < 32\,\UUNIT{MHz}{}\,m_{\rm GeV}^2\, (B/{\mu{\rm
G}})$ and zero above. $\Phi_{10}$ is the original photon flux in
$10^{-10}\,\UUNIT{cm}{-2}\UUNIT{s}{-1}$, $f_{B} = (B^{3/2} {\mu{\rm
G}}^{1/2})/(B^2 + 8\pi\epsilon_{\rm ph}))$ with the photon field
energy density $\epsilon_{\rm ph}$. In the galactic centre 
$\epsilon_{\rm ph}$ can be ignored. In clusters $\epsilon_{\rm ph}
\approx \epsilon_{\rm CMB}$; T.N. LaRosa \textit{et al.}, ApJ.
119, 207 (2000); K.-T. Kim \textit{et al.}, ApJ. 355, 29 (1990). 

\bibitem{sneutrino} J. Hagelin \etal, Nucl. Phys. B241,  638 
(1994). 

\bibitem{heat} HEAT Collaboration,  ApJ. Lett. 482, L191 (1997).

\end{thebibliography}
\end{document}